\documentclass[
 aip, amsmath,amssymb,
reprint, reprint
]{revtex4-2}

\usepackage{graphicx}
\usepackage{bm}
\usepackage{amsmath,stackengine,scalerel,calrsfs}
\usepackage[hidelinks, linkcolor=blue, citecolor=blue, colorlinks=true]{hyperref}
\usepackage{float}
\usepackage[utf8]{inputenc}
\usepackage[T1]{fontenc}
\usepackage{mathptmx}
\usepackage{etoolbox}
\usepackage{enumitem}
\DeclareMathAlphabet{\pazocal}{OMS}{zplm}{m}{n}
\stackMath
\newcommand\reallywidehat[1]{%
\savestack{\tmpbox}{\stretchto{%
  \scaleto{%
    \scalerel*[\widthof{\ensuremath{#1}}]{\kern-.6pt\bigwedge\kern-.6pt}%
    {\rule[-\textheight/2]{1ex}{\textheight}}
  }{\textheight}%
}{0.5ex}}%
\stackon[1pt]{#1}{\tmpbox}%
}
\makeatletter
\def\@email#1#2{%
 \endgroup
 \patchcmd{\titleblock@produce}
  {\frontmatter@RRAPformat}
  {\frontmatter@RRAPformat{\produce@RRAP{*#1\href{mailto:#2}{#2}}}\frontmatter@RRAPformat}
  {}{}
}%
\makeatother
\newcommand{\bA}{\textbf{A}}
\newcommand{\bB}{\textbf{B}}
\newcommand{\bP}{\textbf{P}}
\newcommand{\bq}{\textbf{Q}}
\newcommand{\bE}{\textbf{E}}

\newcommand{\bb}{\textit{\textbf{b}}}

\newcommand{\bJ}{\textbf{J}}
\newcommand{\bj}{\textit{\textbf{j}}}
\newcommand{\bk}{\textit{\textbf{k}}}

\newcommand{\bu}{\textit{\textbf{u}}}
\newcommand{\bx}{\textit{\textbf{x}}}

\newcommand{\bw}{{\boldsymbol{\omega}}}

\newcommand{\bnabla}{{\boldsymbol{\nabla}}}

\begin{document}
\preprint{AIP/123-QED}
\title{Universal energy cascade and relaxation in three-dimensional inertial electron magnetohydrodynamic turbulence}
\author{Supratik Banerjee}%
\email{sbanerjee@iitk.ac.in}
\affiliation{Department of Physics, Indian Institute of Technology Kanpur, Kanpur, Uttar Pradesh, 208016, India}
\author{Arijit Halder}
\affiliation{Department of Physics, Indian Institute of Technology Kanpur, Kanpur, Uttar Pradesh, 208016, India}
\author{Amita Das}
\affiliation{Department of Physics, Indian Institute of Technology Delhi, Hauz Khas, New Delhi 110016, India
}

\date{\today}
\begin{abstract}
Electron magnetohydrodynamics (EMHD) provides a realistic model for electron-scale heating and acceleration in weakly collisional space plasmas. A divergence-free Banerjee-Galtier type (Banerjee and Galtier, JoPA, 2017) exact relation is derived for three-dimensional homogeneous and not necessarily isotropic EMHD turbulence. By explicit calculation, it has been shown that the energy cascade is not affected by the presence of a uniform background magnetic field $\bB_0$. Using direct numerical simulations, a Kolmogorov-like energy cascade with a constant flux rate is observed across the electron inertial scale $d_e$. However, as expected, for length scales greater than $d_e$, a magnetic power spectra of $k^{-7/3}$ is obtained whereas for scales smaller than $d_e$, a $k^{-5/3}$ spectra is obtained. Similar universal cascade rate is also calculated from the scale-by-scale budget in Fourier space and is found to be equal to the one calculated using the exact law in real space. Finally, quenching the turbulence drive, the relaxation of a fully-developed EMHD turbulence is studied using the recently proposed principle of vanishing nonlinear transfers (Banerjee, Halder and Pan, PRE(L), 2023) which convincingly shows the existence of a pressure-balanced relaxed state. 
\end{abstract}

\maketitle

\section{Introduction}
Plasma turbulence is crucial for understanding various aspects of space and laboratory plasmas \citep{Yamada_2008, Banerjee_2013, Bruno_2013, Xu_2017}. Using kinetic and fluid models of plasma, it is possible to tackle a wide range of problems including the energy transport and heating of the solar wind and magnetospheric plasmas \cite{Sorriso_Valvo_2007, Hadid_2018, Horbury_2008, Franci_2015, Sahraoui_2020, Horbury_2012}, the propagation of cosmic rays and highly energetic particles in space \cite{Xu_2020}, energy and angular momentum transport in accretion disks, galaxy clusters and interstellar media \cite{Armitage1998, Lesur2011}, the energy loss in magnetically confined fusion plasma, regulation of heat and particle transport in a fusion chamber for maintaining a sustained fusion, the stability and efficiency of different plasma based industrial applications such as plasma manufacturing and etching \textit{etc.} \cite{Howard2021, White2024}. Recently, turbulence in laser-plasma interactions has garnered significant attention owing to its direct relevance to the challenge of energetic electron stopping in the fast ignition scheme of laser-driven fusion \cite{Chatterjee_2017, Mondal_2012, Mandal_2020, Yabuuchi_2009}. 

The inherent complexity of the plasma medium, comprising at least two particle species with widely differing masses, response times, and length scales, has led to the development of various approximate models to describe its dynamics.
When fluctuation length scales are significantly larger than the ion inertial length $d_i$, the plasma is most accurately described by the magnetohydrodynamic (MHD) fluid model where the momentum of the MHD fluid is mainly provided by the ions \cite{Bittencourt_2013, Banerjee_2020}. This model effectively describes a wide range of phenomena, including the solar wind turbulence beyond the ion inertial scales, the formation of the solar flares, magnetic confinement in fusion experiments and various astrophysical plasmas \cite{Veltri_1999, Shibata_2011, Lao_2021, Ishikawa_1995,  Banerjee2016c, Podesta_2009}. 
However, in many instances, this model falls short of capturing the underlying physical phenomena. For instance, the rapid release of enormous energy during solar flares and the subsequent production of high energetic particles needs a plasma model beyond ordinary MHD remains poorly understood \cite{Gordovskyy_2019, Lenters_1998}. In the framework of solar wind turbulence, the energy power spectra  cannot either be predicted by the model of ordinary MHD turbulence at scales smaller than the ion inertial scales and ion gyroscales \cite{Alexandrova_2009, Sahraoui2009, Chen_2010}. To study plasma turbulence at comparatively smaller scales across $d_i$, Hall MHD model is often used \cite{Banerjee_2016a, Halder_2023, Mininni_2007, Gomez_2010, Banerjee_2024}. However, this model is also found to be insufficient to capture the dynamics around the electron inertial scale.   


An alternative fluid model, tailored to capture fast time-scale phenomena associated with electron dynamics, has been proposed in the literature and is applicable in many scenarios where the MHD and the Hall MHD models are found to be inadequate. This framework is provided by the model of electron magnetohydrodynamics (EMHD), where the ion fluid is at rest, serving solely as a neutralizing background \cite{Kingsep1990, Gordeev1994, Das_1999, Das_2000, Das_2001}. For example, EMHD applies to laser–plasma interaction scenarios only after the laser pulse has exited the plasma, and during subsequent time intervals in which the ions remain effectively immobile.
 As a result, the current density (${\bf J}$) is directly related to the electron fluid velocity ($\bu_e$) through the relation ${\bf J} = - n e \bu_e $ where $n$ denotes the electron number density and  $e$ represents electronic charge. At time scales for which the field propagation speed is much less than the light speed $c$, the displacement current  $({\partial \bf E}/{\partial t})$ can be neglected and the system satisfies $\bnabla\times\bB = - \mu_0 ne \bu_e$ where $\bB$ is the magnetic field.  Interestingly, this assumption is consistent with an incompressible EMHD fluid as ${\boldsymbol \nabla} \cdot {\bu}_e =0 $ .


In the context of space plasmas, EMHD-like models that capture fast, electron-dominated dynamics are frequently invoked. For instance, such models are often employed to describe magnetic reconnection — a process that efficiently converts magnetic energy into kinetic or thermal energy within the thin current sheets observed in solar flares and magnetospheric substorms \citep{Bulanov1992, Stenzel2003}. The turbulent energy cascade characteristic of EMHD can enhance the reconnection rate, resulting in the production of high-energy particles, which are commonly observed in these astrophysical events. Moreover, in collisionless space plasmas, the observed rates of electron-scale heating cannot be accounted for by standard MHD turbulence, which dissipates energy primarily at the respective viscous scales. In contrast, EMHD turbulence naturally leads to the formation of small-scale structures such as current sheets and magnetic islands, which enable efficient dissipation of turbulent energy at electron scales — such as the electron gyroradius ($\rho_e$) and electron inertial length ($d_e$) — through mechanisms like electron Landau damping and magnetic reconnection. In addition, EMHD turbulence can successfully explain electron-scale energy spectra, and to some extent the anomalous heating and acceleration of the solar wind \citep{Alexandrova_2009, Sahraoui2009}. Beyond the solar wind, EMHD instabilities and turbulence can enhance the transport of thermal energy and momentum in accretion disks, and contribute to the generation and amplification of magnetic fields in astrophysical environments via the turbulent dynamo mechanism \cite{LAKHIN_2000, Lakhin_2004}. The evidence of such mechanisms is also found in laboratory laser plasma experiments \cite{Choudhary_2025}.

Due to the intricate nature of complete EMHD model, in practice, the electron inertia is often neglected and the simplified inertialess EMHD model is employed to investigate electron-scale turbulence in plasmas.
A considerable amount of research has been conducted investigating several aspects of inertialess EMHD turbulence, including magnetic energy and magnetic helicity cascades, power spectra, anisotropy, and wave turbulence. Despite remarkable simplicity, this model successfully produces an isotropic $k^{-7/3}$ power-spectra for magnetic energy while an anisotropic  $k_{\perp}^{-8/3}$ power spectra is obtained for negligible parallel transfer \cite{Cho_2004, Cho_2009, Meyrand_2013}. Similar to ordinary MHD, an inverse cascade of magnetic helicity is also observed from smaller to the larger scales \cite{Cho_2011, Banerjee_2016a, Kim_2015, Cho_2016a, Cho_2016b}. The wave turbulence of inertialess EMHD is essentially mediated by the electron whistler waves where the weak cascades of magnetic energy and magnetic helicity correspond to anisotropic $k_{\perp}^{-5/2} k_{\parallel}^{-1/2}$ and $k_{\perp}^{-7/2} k_{\parallel}^{-1/2}$ spectra, respectively \cite{Galtier_2003, Galtier_2005}.       

However, the inertialess EMHD model becomes inadequate for analyzing turbulent behavior at scales smaller than $d_e$. Therefore, a systematic investigation of the more general (inertial) EMHD turbulence is essential for a comprehensive understanding of plasma turbulence across $d_e$. In contrast to the relatively well-studied inertialess EMHD turbulence, only a limited number of studies have focused on EMHD turbulence that includes electron inertia effects.
Using direct numerical simulations in two and three-dimensions, it has been shown that the small-scale turbulence in EMHD is generated through Kelvin-Helmholtz instability where the energy dissipation rates are found to be independent of diffusion coefficients. Moreover, in the spectral space, a whistler driven $k^{-7/3}$ and Kolmogorov-like $k^{-5/3}$ spectra for the total energy are reported for length scales larger and smaller than $d_e$, respectively \cite{Biskamp_1999, Biskamp_2003}. For strong EMHD turbulence, a von K\'{a}rm\'{a}n-Monin type exact relation was derived to study the corresponding energy cascade using two-point statistics but it was numerically studied for two-dimensional electron MHD turbulence only \cite{Celani_1998a, Celani_1998b}. Using the principle of selective decay, double-curl Beltrami relaxed states are predicted for an incompressible EMHD flow \cite{Bhattacharyya_2003}. In addition, some aspects of intermittency in 2D EMHD turbulence have also been explored \cite{Boffetta_1998, Shivamoggi_2015}. Recently, the wave turbulence in EMHD has also been investigated \cite{David_2022}. Despite these works, a dedicated analytical and numerical study on the energy cascade in three dimensional EMHD turbulence is still lacking. 

In this paper, we derive a divergence-free Banerjee-Galtier type (BG17) exact relation \cite{Banerjee_2016a, Banerjee_2016b} for the total energy cascade of three dimensional fully developed EMHD turbulence without neglecting the electron inertia. We then investigate the relevant large scale and small scale limits and the influence of a uniform background magnetic field on the cascade rates. The obtained exact law is expressible in terms of the two-point increments and hence can be useful to predict the scaling properties of different quantities. Using recently proposed principle of vanishing non-linear transfer (PVNLT) \cite{Banerjee_2023}, we also predict the turbulent relaxed states which are more general than those obtained using principle of selective decay. Finally, with the aid of direct numerical simulations with $256^3$ and $512^3$ grid points, we numerically search for a Kolmogorov-like universal energy cascade and calculate the corresponding cascade rate which effectively provides the respective turbulent heating rate in three-dimensional EMHD turbulence. Finally, we quench the turbulence forcing to obtain the corresponding relaxed states and characterize them. 

The paper is organized as follows: the governing equations and the inviscid invariants of EMHD are presented in sec.~\ref{sec2A} whereas the sec.~\ref{sec2B} mainly consists of the derivation of the exact relation for fully developed EMHD turbulence. The subsequent sections \ref{sec2C} and \ref{sec2D} are dedicated to the reduction of the exact relation at different limits and to study the effect of a uniform background magnetic field, respectively. In the next section, using PVNLT, we predict the turbulent relaxed states which is followed by sec. \ref{sec2F} where the cascade rates are derived in terms of flux rates in Fourier space. The entire section \ref{sec3} provides a step-by-step description of numerical simulations. The three subsections \ref{sec3A}, \ref{sec3B} and \ref{sec3C} systematically describe the non-dimensionalisation of the governing equations, the numerical set-up and an elaborate presentation of the numerical findings, respectively. Finally, in sec. \ref{sec4}, we summarize, highlight the significance of the results and discuss some possible prospective projects as an extension of the present work. 

\section{Theoretical framework}
\label{sec2}
\subsection{Governing equations and inviscid invariants}
\label{sec2A}
The main equations of incompressible EMHD consist of the evolution equation for the momentum of the electron fluid and the induction equation and can be written as 
\begin{align}
   \frac{\partial{\bu_e}}{\partial t} + ({\bu_e} \cdot \bnabla ) {\bu_e}&=  - \frac{\bnabla p_e}{\rho_e} - \frac{e}{m_e} ({\bf E} + {\bu_e \times \bB}) + \bm{f} + \bm{D}_e, \label{emhd_mom} \\
    \frac{\partial{\bB}}{\partial t} &= - \bnabla \times {\bf E}\label{emhd_Faraday},
     \end{align}
where $\rho_e$ is the constant density of the electron fluid, $p_e$ is the fluid pressure, $\bE$ is the electric field, $\bm{f}$ is the external forcing and $\bm{D}_e$ denotes the dissipation term where 
\begin{equation}
  \bm{D}_e= -\nu_{ei} \bu_e + \nu_e \nabla^2 \bu_e,
\end{equation}
with $\nu_{ei}$ being the ion-electron collision frequency and $\nu_e$ representing the kinematic viscosity of the electron fluid. The total energy density can be given by the sum of the densities for electron fluid kinetic energy and magnetic energy as
\begin{equation}
\pazocal{E} = \frac{1}{2} \rho_e u_e^2 + \frac{B^2}{2 \mu_0}.\label{energy_density_v1}
\end{equation}
Using the governing equations \eqref{emhd_mom} and \eqref{emhd_Faraday} one can write 
\begin{align}
\frac{\partial}{\partial t} &\left(\frac{1}{2} \rho_e u_e^2 + \frac{B^2}{2 \mu_0} \right)\nonumber\\
= &\left( \rho_e \bu_e \cdot  \frac{\partial \bu_e}{\partial t} + \frac{1}{ \mu_0} \bB \cdot \frac{\partial \bB}{\partial t} \right) \nonumber\\
=& - \bnabla \cdot \left[\left( \frac{1}{2} \rho_e u_e^2 + p_e \right) \bu_e +\frac{1}{\mu_0}\bE \times \bB\right]\nonumber\\
&+ \rho_e\bu_e\cdot\bm{f} + \rho_e\bu_e\cdot\bm{D}_e. \label{Etot_evolution}
    \end{align}
Therefore, it is straightforward to write  
\begin{equation}
    \frac{d \langle \pazocal{E}\rangle}{d t} = \varepsilon^{diss} + \varepsilon  , \label{Etot_conv} 
\end{equation}
where $\varepsilon^{diss} = -\langle\nu_{ei} \rho_e u_e^2 + \nu_e \rho_e {\omega}_e^2\rangle $, $\varepsilon = \langle\rho_e\bu_e\cdot\bm{f}\rangle$, $\bw_e = \bnabla\times\bu_e$, $\langle(\cdot)\rangle$ denotes the statistical average (equivalent to space average) and the divergence term vanishes under periodic or vanishing boundary conditions. From analogy with ordinary MHD,  the term $\nu_e \rho_e {\omega}_e^2$ 
is expected to correspond the dissipation of electron fluid kinetic energy whereas the term $\nu_{ei} \rho_e u_e^2$ represents the dissipation of magnetic energy as $\nu_{ei} \rho_e u_e^2  = \eta \bJ^2$ where $\eta= m_e \nu_{ei}/\mu_0 ne^2$ is the magnetic diffusivity. From Eq.~\eqref{Etot_conv}, it is straightforward to understand that $\langle \pazocal{E}\rangle$ remains constant over time if a mutual balance between $\varepsilon^{diss}$ and $\varepsilon$ is achieved.


In electron MHD, both the momentum evolution and induction equations are coupled through the electric field $\bE$ and therefore not possible to separately write the inviscid evolution equations for $\bu_e$ and $\bB$. However, elimination of $\bE$ from Eqs.~\eqref{emhd_mom} and \eqref{emhd_Faraday} leads to
\begin{equation}
   \frac{\partial {\bf Q}}{\partial t} = \bnabla \times \left( \bu_e \times {\bf Q} \right) + \bm{f}_{\bq} +\bm{D}_{\bf Q}, \label{frozen_in_Eq}
\end{equation}
where ${\bf Q} =\bB - (m_e\bw_e)/e= \bB - d_e^2 \nabla^2 \bB$, $\bm{f}_{\bq} = -(m_e/e)\bnabla\times\bm{f}$ and $\bm{D}_{\bf Q}= \eta\nabla^2 \bB - \nu_e d_e^2\nabla^2\nabla^2\bB$ represents the dissipative term and $d_e = c/\omega_{pe}= \left(m_e/\mu_0 n e^2\right)^{1/2}$ is the electron inertial length. For larger scales, where $k d_e << 1$, the field ${\bf Q}$ is effectively reduced to magnetic field whereas for smaller scales, where $k d_e >> 1$, the variable effectively represents the electron fluid vorticity $\bw_e$. As it will be discussed later, for numerically studying the energy cascade in EMHD turbulence, rather than simulating 
Eqs.~\eqref{emhd_mom} and \eqref{emhd_Faraday}, we simulate Eq.~\eqref{frozen_in_Eq}.

Besides total energy, the canonical helicity $\langle H\rangle = \langle\bP \cdot {\bf Q}\rangle$, where ${\boldsymbol \nabla} \times {\bf P} = {\bf Q} $, is also an inviscid invariant of incompressible EMHD. The conservation and the corresponding exact relation are discussed in Appendix \ref{appA}. However, a separate study is required to numerically explore different aspects of a stationary helicity cascade in EMHD turbulence and will be prepared in future.

\subsection{The derivation of the exact relation}
\label{sec2B}
In this section, we derive an exact relation for the inertial range transfer of total energy in three-dimensional EMHD turbulence. First we construct the symmetric two-point correlator for the total energy as 
\begin{equation}
    R_{\pazocal{E}} = \left\langle \frac{\rho_e}{2}  \bu_e \cdot \bu'_e + \frac{\bB \cdot \bB'}{2 \mu_0}\right\rangle , 
\end{equation}
where the unprimed and primed quantities represent field properties at points $\bx$ and $\bx+\bm{\ell}$ respectively. The corresponding evolution of the correlators is given by
\begin{align}
    \frac{\partial R_{\pazocal{E}}}{ \partial t} 
    &= \left\langle \frac{\rho_e}{2} \frac{\partial \bu_e}{ \partial t} \cdot \bu'_e + \frac{\rho_e}{2}  \bu_e \cdot \frac{\partial \bu'_e}{ \partial t} + \frac{\bB' }{2 \mu_0} \cdot \frac{\partial \bB}{ \partial t} + \frac{\bB }{2 \mu_0} \cdot \frac{\partial \bB'}{ \partial t} \right\rangle \nonumber \\
    &=\left\langle \frac{\rho_e}{2}  \bu'_e \cdot \left[ - ( \bu_e \cdot \bnabla) \bu_e - \frac{\bnabla p_e}{\rho_e} - \frac{e}{m_e} \left( \bE + \bu_e \times \bB \right) \right]   \nonumber \right. \\
    &\left.+ \frac{\rho_e}{2}  \bu_e \cdot \left[ - ( \bu'_e \cdot \bnabla') \bu'_e - \frac{\bnabla' p'_e}{\rho_e} - \frac{e}{m_e} \left( \bE' + \bu'_e \times \bB' \right)  \right]   \nonumber \right. \\
     &\left.-\frac{\bB' }{2 \mu_0} \cdot ( \bnabla \times \bE ) - \frac{\bB }{2 \mu_0} \cdot ( \bnabla' \times \bE') \right\rangle + {\cal D} + {\cal F},\label{time_derivative_correlator}
\end{align}
where $\mathcal{D} = \left\langle \nu_e\rho_e\bu_e\cdot\nabla^{'2}\bu'_e + \nu_e\rho_e\bu'_e\cdot\nabla^{2}\bu_e -2\nu_{ei}\rho_e\bu_e\cdot\bu'_e  \right\rangle/2$ and $\mathcal{F} = \left\langle\rho_e\bu_e\cdot\bm{f}'+\rho_e\bu'_e\cdot\bm{f}\right\rangle/2$
represent the two-point dissipative and forcing contributions, respectively. Using the properties of homogeneous turbulence, one can write 
\begin{equation}
\left\langle  -\bB' \cdot ( \bnabla \times \bE ) \right\rangle = \left\langle \bnabla' \cdot ( \bE \times \bB' ) \right\rangle =ne \mu_0  \left\langle \bE \cdot \bu'_e \right\rangle
\end{equation}
and similarly, 
\begin{equation}
    \left\langle  -\bB \cdot ( \bnabla' \times \bE' ) \right\rangle =  n e \mu_0  \left\langle \bE' \cdot \bu_e \right\rangle.
\end{equation}
With these manipulations, one can write 
\begin{align}
\frac{\partial R_\pazocal{E}}{\partial t}\nonumber
 &= \frac{\rho_e}{2}\left\langle -\bu_e'\cdot ( \bu_e \cdot \bnabla) \bu_e-\bu_e\cdot ( \bu_e' \cdot \bnabla') \bu_e'\right\rangle\nonumber\\              
 &+\frac{ne}{2}\left\langle -\bu_e'\cdot\left( \bu_e \times \bB \right)-\bu_e\cdot\left( \bu_e' \times \bB' \right)\right\rangle +{\cal D} + {\cal F},
\label{time_derivative_correlator_simplified}
\end{align}
where the pressure contributions vanish under the assumption of homogeneity and incompressibility as $\bnabla\cdot\bu_e=\bnabla'\cdot\bu_e'=0$.
Using the identity $({\bu_e} \cdot \bnabla ) {\bu_e} = \bnabla ({u_e^2}/{2}) - {\bu_e} \times {\boldsymbol \omega}_e$, it is straightforward to write
\begin{equation}
\frac{\partial R_{\pazocal{E}}}{ \partial t} = \frac{ne}{2} \left\langle  \bu'_e\cdot\left(\bq\times \bu_e \right) +  \bu_e\cdot\left(\bq'\times\bu'_e\right)   \right\rangle + {\cal D} + {\cal F}.\label{time_derivative_correlator_final}
\end{equation}
Now, if we consider a stationary regime for the transfer of energy, the left hand side of the above equation vanishes. For inertial length scales far from the dissipative scales, one can neglect $ {\cal D}$
and finally considering the forcing scales to be far away from the inertial scales 
one can show that $\mathcal{F} = \left\langle\rho_e\bu_e\cdot\bm{f}'+\rho_e\bu'_e\cdot\bm{f}\right\rangle/2 \approx \left\langle\rho_e\bu'_e\cdot\bm{f}'+\rho_e\bu_e\cdot\bm{f}\right\rangle/2 =\langle\rho_e\bu_e\cdot\bm{f}\rangle = \varepsilon$ then the above Eq.~\eqref{time_derivative_correlator_final} reduces to  
\begin{equation}
  A(\bm{\ell}) = ne \left\langle \delta\left(\bq\times\bu_e \right) \cdot \delta \bu_e \right\rangle = 2 \varepsilon.\label{exact_law_energy_final} 
\end{equation}
Eq.~\eqref{exact_law_energy_final} is the main result of this paper. It provides
an exact measure for the turbulent energy transfer across inertial scales in fully  developed homogeneous three-dimensional EMHD turbulence including the electron inertia. 
As it is seen from Eq.~\eqref{exact_law_energy_final}, $\varepsilon$ can be determined from the knowledge of all $\bB$, $\bu_e\ (\propto\bnabla\times\bB)$ and $\bw_e\ (\propto\bnabla\times\bnabla\times\bB)$ whereas no information about electric field $\bE$ is required.  

\subsection{Different limits}
\label{sec2C}
The derived exact relation \eqref{exact_law_energy_final} takes the BG17 form and closely resembles the exact relation derived for the ordinary hydrodynamic turbulence with the fluid velocity and vorticity replaced by the electron fluid velocity $\bu_e$ and generalised vorticity $\bq$, respectively. However, care must be taken while testing different limits of the above exact relation. For example, it is tempting to recover a `hydrodynamic limit' of the electron fluid simply by putting $\bB=\bm{0}$ in Eq.~\eqref{exact_law_energy_final}. However, a zero $\bB$ implies zero $\bJ$, which results in zero $\bu_e$ in the EMHD framework and thus causing the total turbulent energy transfer to vanish. As $\bq = \bB - d_e^2 \nabla^2 \bB $, two different limits can be obtained for length scales much larger and smaller than $d_e$ and the resulting exact relations are given by 
\begin{align}
   - n e \left\langle \delta \left( \bu_e \times \bB \right) \cdot \delta \bu_e \right\rangle &= 2 \varepsilon, \, \, \text{for } k d_e \ll 1 \;\; \text{and}\label{exact_law_HMHD} \\
   \rho_e \left\langle \delta \left( \bu_e \times  \boldsymbol{\omega}_e \right)  \cdot \delta \bu_e \right\rangle  &= 2 \varepsilon, \, \, \text{for } k d_e \gg 1 \label{exact_law_EMHD_fluid}
\end{align}
respectively. Veritably, Eq.~\eqref{exact_law_HMHD} is similar to the exact relation for energy transfer rate in inertia-less EMHD regime \citep{Banerjee_2016b} whereas Eq.~\eqref{exact_law_EMHD_fluid} provides the transfer rate of electron fluid kinetic energy in inertia-dominated EMHD regime.

Starting from the exact relation derived for the total energy cascade of compressible two-fluid turbulence \citep{Banerjee_2020}, the exact relation for energy cascade in EMHD turbulence can be recovered in the appropriate limit. Unlike the center of mass fluid models such as ordinary MHD or HMHD, EMHD model represents a single species fluid and hence the Eq.~\eqref{exact_law_energy_final} can be obtained as an incompressible two-fluid limit with $\bu_i =\bm{0}$ (see the Eq. (31) of Ref.~\citep{Banerjee_2020}). Again, the exact relation for energy is also derived from quasi-neutral two-fluid plasma turbulence in the limit of ${\bu}_i = \bm{0}$ \citep{Andres_2016}. However, note that, the quasi-neutrality is not a necessary condition for two-fluid plasmas and can be justified only if the plasma is highly ionized. 

\subsection{Effect of a uniform background magnetic field}
\label{sec2D}
The presence of a background magnetic field $\bB_0$ entails power and spectral index anisotropy in MHD turbulence. However, the energy cascade rates of MHD and HMHD turbulence are shown to remain unaltered in the presence of $\bB_0$. For EMHD turbulence, it is also interesting to explore the effect of mean magnetic field on the cascade rate. For that, let us decompose the magnetic field $\bB$ as $\bB = \bB_0 + \Tilde{\bB}$, where $\bB_0$ is the mean field and $\Tilde{\bB}$ is the fluctuation. Since, the electron fluid velocity $\bu_e$ is proportional to the current $\bJ = (\bnabla \times \bB)/ \mu_0$, it does not have any contribution from $\bB_0$. The same applies for $\boldsymbol{\omega}_e$ and hence $\bq = \tilde{\bB} - (m_e\bw_e)/e + \bB_0 = \tilde{\bq} + \bB_0$. By straightforward algebra, one can show the contribution from the mean field vanishes as $n e \left\langle \left(\bB_0\times \delta\bu_e \right) \cdot \delta \bu_e \right\rangle = 0$.

\subsection{Turbulent relaxed states}
\label{sec2E}
From Eq.~\eqref{exact_law_energy_final}, it is evident that the condition $\bu_e \parallel \bq$ causes the average nonlinear transfer rate $A(\bm{\ell})$ to vanish. In general, such aligned conditions are associated with the relaxed states of a turbulence system when the forcing is switched off. However, more general relaxed states for EMHD turbulence can be obtained by using the recently proposed PVNLT. In the said framework, we search for the non-static relaxed states ($\bu_e \neq \bm{0}$) where average transfer rate of the turbulent cascade vanishes across all scales in the inertial range. Such a condition is mathematically expressed as 
\begin{equation}
  \bu_e\times\bq = \bnabla \Phi \implies\bnabla\times(\bu_e\times\bq) = \bm{0},\label{relaxed_solution}
\end{equation}
where $\bnabla\Phi$ effectively represents the gradient of an effective pressure $p_e+\rho_e u_e^2/2+\theta$ where $\theta$ is an arbitrary scalar function. Note that, the above condition is also consistent with vanishing of the canonical helicity cascade which can be shown from Eq.~\eqref{time_derivative_correlator_final_H}. In case of negligibly small $\bnabla\Phi$, we have $\left(\bB -m_e\bw_e/e\right) = \beta \bu_e$, where $\beta$ is the proportionality constant. Using the relation between $\bu_e$ and $\bJ$,  this condition can also be expressed as a double-curl Beltrami state (similar to Ref.~\cite{Bhattacharyya_2003})  
\begin{equation}  
  d_e^2 \bnabla \times \left(\bnabla \times \bB \right) + \lambda \left(\bnabla \times \bB \right) + \bB = \bm{0},\label{double_curl_state}
\end{equation}
where $\lambda = \beta/ne\mu_0$. Similar form of relaxed states are also obtained for the turbulent relaxation of an incompressible HMHD flow and obviously includes Taylor-Beltrami alignment ( $\bnabla \times \bB \parallel \bB$) as one of its (but not the only one) exact solutions. We can also look for the relaxed conditions for length scales larger and smaller than $d_e$. In the former case, Eq.~\eqref{relaxed_solution} gives $\bu_e\times\bB = \bnabla\Phi$ which results in a dynamic aligned state $\bu_e\times\bB = \bm{0}$ for negligible $\bnabla\Phi$. Due to the proportionality between $\bu_e$ and $\bJ$ in EMHD, this dynamic aligned state is also equivalent to a Beltrami-Taylor alignment. This is in stark contrast with ordinary MHD where dynamic alignment and the Beltrami-Taylor states are found to be mutually exclusive \cite{Banerjee_2023}. In case where $kd_e\gg 1$, Eq.~\eqref{relaxed_solution} reduces to $\bu_e\times\bw_e = -(e/m_e)\bnabla\Phi$ which gives, similar to hydrodynamic turbulence, a Beltrami alignment between $\bu_e$ and $\bw_e$ when the pressure gradient is neglected. Again for EMHD, this is equivalent to an alignment between $\bJ$ and $\bnabla\times\bJ$. 

Although the Eq.~\eqref{exact_law_energy_final} provides exact analytical expressions for the cascade rate $\varepsilon$, it remains unclear whether it is constant across the inertial range. Upon quenching the turbulent forcing, it is also not evident whether such a flow relaxes towards an aligned state or a state supporting a non-zero pressure-gradient. To conduct a conclusive study about the universality of the cascades and the relaxed states, we therefore perform direct numerical simulations of a three-dimensional inertial EMHD flow and analyze the outcome by means of the theoretical framework outlined above. 

\subsection{Cascade rate in the Fourier space}
\label{sec2F}
Unlike the previous method, where $\varepsilon$ was calculated in terms of two-point increments, one can equivalently obtain the cascade rates by explicitly calculating the energy flux rates in Fourier space. By the help of Eqs.~\eqref{emhd_mom} and \eqref{emhd_Faraday}, one can calculate the evolution equations for $\pazocal{E}_K (\bk)\ =\rho_e\bu_e (\bk)\cdot\bu^*_e (\bk)/2 $ and $\pazocal{E}_M (\bk)\ =\bB(\bk)\cdot\bB^* (\bk)/2\mu_0$ and show that
\begin{equation}
    \frac{\partial}{\partial t} (\pazocal{E}_K + \pazocal{E}_M)= A(\bk) + {\cal D}(\bk) + {\cal F}(\bk),\label{spectra_scale_balance}
\end{equation}
where 
\begin{align}
  A(\bk) &= \Re\left[\rho_e\bu_e^* (\bk)\cdot\widehat{(\bu_e\times\bw_e)}_{\bk}-ne\ \bu_e^* (\bk)\cdot\widehat{(\bu_e\times\bB)}_{\bk}\right]\nonumber\\
  &=ne\Re\left[\ \bu_e^* (\bk)\cdot\widehat{\left(\bq\times\bu_e\right)}_{\bk}\right]
\end{align}
is the net nonlinear energy transfer rate corresponding to the $\bk$-th mode, ${\cal D}(\bk) = \Re\left[\rho_e \bu_e^* (\bk)\cdot\bm{D}_e (\bk)\right]$ and  ${\cal F}(\bk) = \Re\left[\rho_e \bu_e^* (\bk)\cdot\bm{f}(\bk)\right]$ with $\Re$ and $\widehat{(\cdot)}$ denoting the real part of a complex number and Fourier transform, respectively. For stationary turbulence, the \textit{l.h.s.} of Eq.~\eqref{spectra_scale_balance} vanishes. Neglecting the dissipative contribution well inside the inertial zone, the Eq.~\eqref{spectra_scale_balance} reduces to 
\begin{equation}
   A(\bk) +  {\cal F}(\bk) = 0.
\end{equation}
For large-scale forcing acting at $k_f$, one can write 
\begin{equation}
    \sum_{|\bk'|\leq k}\Re\left[\rho_e \bu_e^{*} (\bk')\cdot\bm{f}(\bk')\right] \approx \sum_{|\bk'| = k_f}\Re\left[\rho_e \bu_e^{*} (\bk')\cdot\bm{f}(\bk')\right] = \varepsilon.
\end{equation}
Finally, the average energy flux rate across a sphere of radius $k$ is given by 
\begin{equation}
\Pi (k)= -ne\sum_{|\bk'|\leq k}\Re\left[\ \bu_e^* (\bk')\cdot\widehat{\left(\bq\times\bu_e\right)}_{\bk'}\right] = \varepsilon. \label{Flux_rate_energy_Fourier} 
\end{equation}

\section{Numerical investigation}
\label{sec3}
\subsection{Non-dimensionalisation of governing equations}
\label{sec3A}
For numerical analysis, we take $\rho_e = 1$ and rewrite the magnetic field in the Alfv\'{e}n units such that $\bb= \bB/\sqrt{\mu_0}$, $\bj = \bnabla\times\bb$, $\bu_e = -d_e\bj$, $\bw_e = d_e\nabla^2 \bb$ and $\bm{q}=\bb - d_e\bw_e$. In terms of the newly introduced variables, along with a proper non-dimensonalisation, Eq.~\eqref{frozen_in_Eq} can be re-written as 
\begin{equation}
\frac{\partial \bm{q}^*}{\partial t^*}
=\bnabla^*\times\left(\bu_e^*\times\bm{q}^*\right) +\bm{f}^*_{\bm{q}^*} + \eta^* \nabla^{*2}\bb^* - \nu_e^* d_e^{*2} \nabla^{*4}\bb^*,\label{Geq_nondim}
\end{equation}
where $\bm{q}^* = \bb^*-d_e^{*2}\nabla^{*2}\bb^*$, $\bb^* = \bb/U_{rms}$, $\bu_e^* = \bu_e/U_{rms}$, $\bm{f}^*_{\bm{q}^*}= (L/U^2_{rms}) \bm{f}_{\bm{q}}$, $ t^*= (U_{rms}/L)t$, $\bnabla^* = L\bnabla$, $d_e^* =  d_e/L$, $\eta^*=\eta/(U_{rms} L) $ and $\nu_e^*=\nu_e/(U_{rms} L) $ with $L$ being the box size. For the sake of notational simplicity, we drop the * sign hereinafter. In our numerical simulations, for convenience, we choose $\nu_e = \eta$ such that Eq.~\eqref{Geq_nondim} reduces to 
\begin{equation}
    \frac{\partial \bm{q}}{\partial t} = \bnabla\times(\bu_e\times\bm{q}) +\bm{f}_{\bm{q}} + \eta \nabla^2 \bm{q}.\label{gov_eq_nondim_final}
\end{equation}
In a similar manner as above, the derived exact relation in Eq.~\eqref{exact_law_energy_final} can be written in the non-dimensional units as 
\begin{equation}
A(\bm{\ell}) = d_e^{-1}\left\langle \delta\left(\bm{q}\times\bu_e \right) \cdot \delta \bu_e \right\rangle = 2 \varepsilon \label{exact_energy_nondim}\\
\end{equation}
whereas the flux rate in Eq.~\eqref{Flux_rate_energy_Fourier} and is given by 
\begin{equation}
 \Pi (k) = - d_e^{-1}\sum_{|\bk'|\leq k}\Re\left[\bu_e^* (\bk')\cdot\widehat{\left(\bm{q}\times\bu_e\right)}_{\bk'}\right] =\varepsilon,\label{flux_Fourier_energy_nondim}
\end{equation}
Finally, the relaxed state in Eq.~\eqref{relaxed_solution} is written in the non-dimensional units as
\begin{equation}
    \bu_e\times\bm{q}=\bnabla\Phi\implies \bnabla\times(\bu_e\times\bm{q}) = \bm{0}.\label{relaxed_state_nondim}
\end{equation}

\subsection{Numerical set-up}
\label{sec3B}
\begin{figure}
    \centering
    \includegraphics[width=0.9\columnwidth]{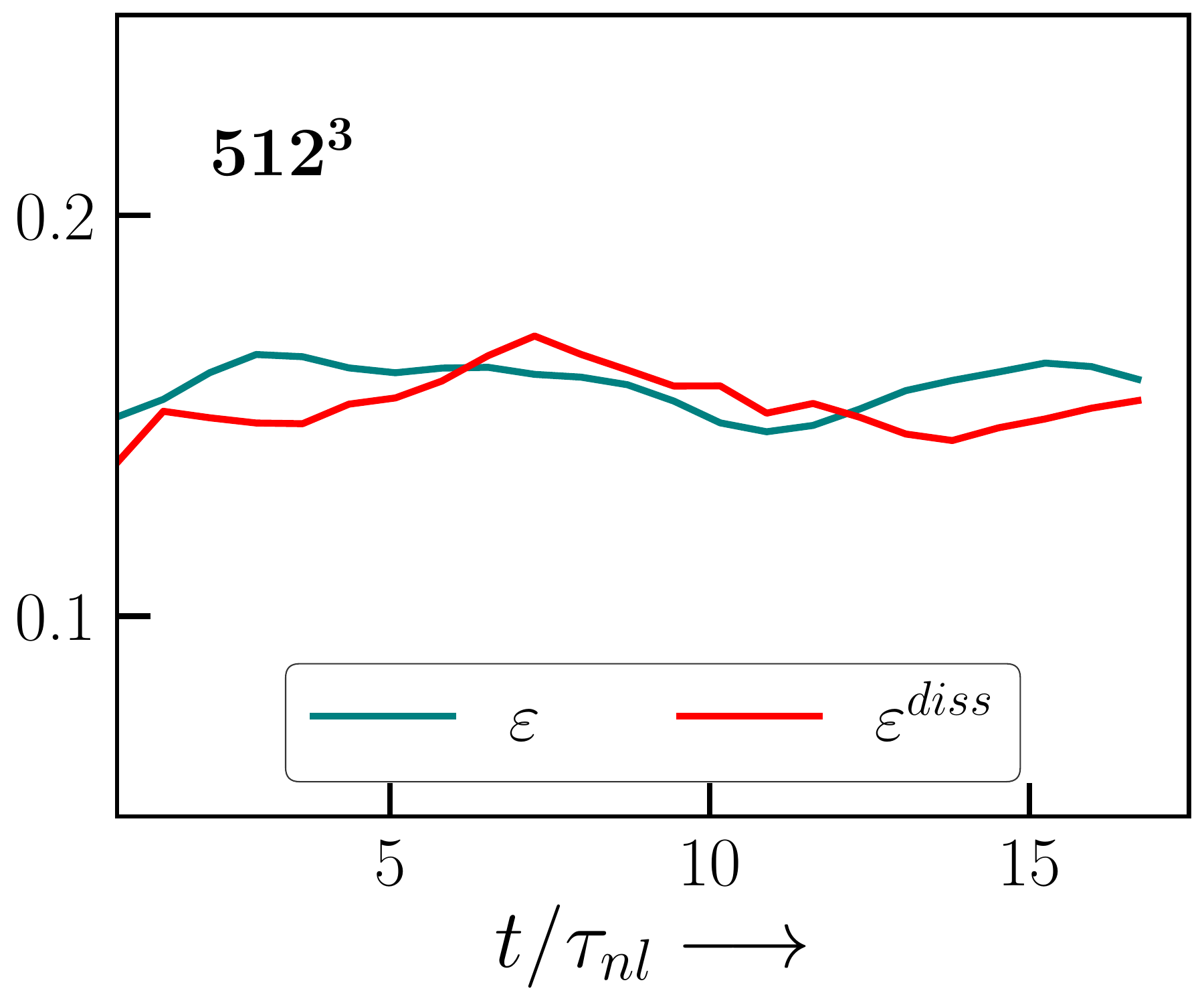}
    \caption{Average injection and dissipation rates for energy as function of time (normalised by integral-scale nonlinear time $\tau_{nl}$). For larger times, a fair balance between average injection and dissipation rates is achieved thus entailing stationarity for $\langle \pazocal{E}\rangle$. }
    \label{stationary_state}
\end{figure}
\setlength{\tabcolsep}{0.5em} 
{\renewcommand{\arraystretch}{1.1}
\begin{table}
\centering
\begin{tabular}{l l l l l l l l}
\hline\hline
Run &  $N$&  $\eta$& $d_e$& $L$& $\ell_d$& $k_{max}$& $k_{max}/k_{d}$   \\
\hline
A &  $256$&  $5\times 10^{-7}$& $0.05$& $0.71$& $0.098$& $85$& $1.32$ \\
B &  $512$&  $5\times 10^{-8}$& $0.05$& $0.72$& $0.049$& $171$& $1.32$  \\
\hline\hline
\end{tabular}
\caption{Simulation parameters for all the runs:  $N^3$ denotes the total number of grid points, $L = (3\pi/4)\int E(k)k^{-1}\ dk/\int E(k)\ dk$ is the integral length scale where $E(k)$ is the spectra for total energy, $\ell_d= 2\pi/k_d$ is the Kolmogorov length scale where $k_d = (\varepsilon/\eta^3)^{0.1}$ with $\varepsilon$ being the total dissipation rate and $\tau_{nl} = \ell/\sqrt{\langle(\delta u_{e\parallel}^2(\ell)\rangle}$ is the integral-scale nonlinear time. }
\label{table_parameters}
\end{table}}
We solve Eq.~\eqref{gov_eq_nondim_final} in a 3D periodic domain of size $2\pi$ using a MPI-based parallel pseudo-spectral code written in Python. For time integration, we employ standard fourth-order Runge-Kutta (RK4) method. The aliasing error is minimised with 2/3-rd de-aliasing method leading to maximum available wavenumber $\approx N/3$ where $N$ is the number of grid points in each spatial direction. Initializing $\bm{q}$ with a random condition,  the turbulence is generated by adding a large-scale non-helical Taylor-Green forcing $\bm{f}_{\bm{q}}= f_0 [sin(k_0x)cos(k_0y)cos(k_0z), -cos(k_0x)sin(k_0y)cos(k_0z), 0]$ with $f_0 = 0.5$ and $k_0=2$. 
To get an uncontaminated inertial range, we use a fourth order $(\sim \nabla^4)$ hyperdiffusive operator for $\bm{q}$. We perform numerical simulations with both $256^3$ and $512^3$ grid points with input Reynolds numbers $2\times 10^6$ and $2\times 10^7$, respectively. For both the runs, the nondimensional electron inertial length scale is taken to be 0.05 and the simulations are well resolved as for the both runs $k_{max}/k_d > 1$ where $k_d =2\pi/\ell_d$ with $\ell_d$ being the Kolmogorov length scale. However, as expected, the $k_{max}$ for the DNS with $512^3$ grid points is twice the $k_{max}$ of DNS with $256^3$ grid points, thus leading to an increased inertial range. The relevant simulation parameters are summarized in Tab.~\ref{table_parameters}.

\subsection{Results}
\label{sec3C}
For forced turbulence, the inertial range cascade rates are calculated from the exact relations at a statistical stationary regime achieved through a balance between average injection and dissipation rates of the cascading invariants. To ascertain such a regime, we run the simulations for several nonlinear times $(\tau_{nl})$ where $\tau_{nl} = \ell/\sqrt{\langle(\delta u_{e\parallel}^2(\ell)\rangle}$. In particular, the total run times are approximately $62\tau_{nl}$ and $16\tau_{nl}$ for simulations with $256^3$ and $512^3$ grid points, respectively. For the sake of clear visualization, we over-plot
the average injection and dissipation rates for the total energy in Fig.~\ref{stationary_state} as a function of time (normalised by $\tau_{nl}$) for run B. A reasonable balance between the average injection and dissipation rates is observed for larger times signifying stationarity of $\langle\pazocal{E}\rangle$.

\begin{figure}
    \centering
    \includegraphics[width=\columnwidth]{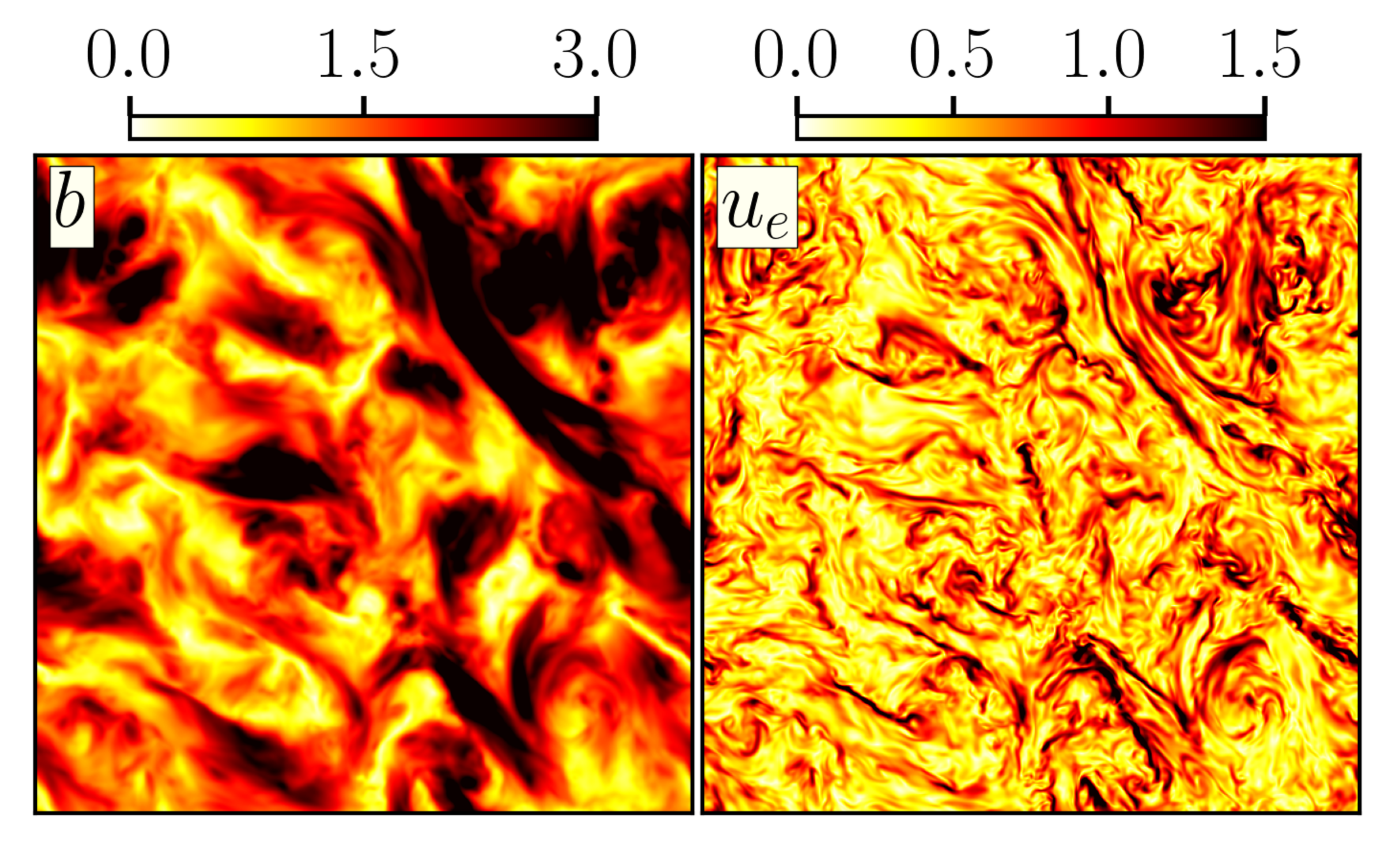}
    \caption{Instantaneous snapshots for $b$ and $u_e$ at $15\tau_{nl}$ show structures at different scales. The magnetic field is dominated by large-scale structures whereas the velocity field show prevalence of small-scale finer structures.  }
    \label{fig:2D_slices}
\end{figure}

In Fig.~\ref{fig:2D_slices}, we plot instantaneous snapshots for $b$ and $u_e$ at $15\tau_{nl}$ for the simulation with $512^3$ grid points. Both snapshots consist of structures at different length-scales signifying a state of fully developed homogeneous turbulence. As it is apparent, large-scale structures are prevailing for $b$ while $u_e$ is dominated by finer small-scale structures. This subsequently indicates the dominance of magnetic energy at large-scales in contrast with the electron fluid kinetic energy overpowering at small-scales. 

To compute the relevant structure functions in Eq.~\eqref{exact_energy_nondim}, we employ a method based on the isotropic SO(3) decomposition \cite{Taylor_2003}. Here, the two-point differences are calculated by varying the increments in 73 directions covered by the base vectors (in the units of grid separation) $\in \{(1, 0, 0),\ (1, 1, 0),\ (1, 1, 1),\ (2, 1, 0),\ (2, 1, 1),\ (2, 2, 1),$ $\ (3, 1, 0),\ (3, 1, 1)\}$ and those generated by the all possible permutations and sign changes. This ensures that the simulation domain is homogeneously spanned in all the directions without any interpolation of data points. Finally, the structure functions calculated for all 73 directions are averaged to obtain the cascade rates as a function of isotropic scale $\ell$. However, the unit vectors for all the directions are not the same and hence a one-dimensional interpolation is required for the structure functions before the final averaging. 
\begin{figure}
    \centering
    \includegraphics[width=0.9\columnwidth]{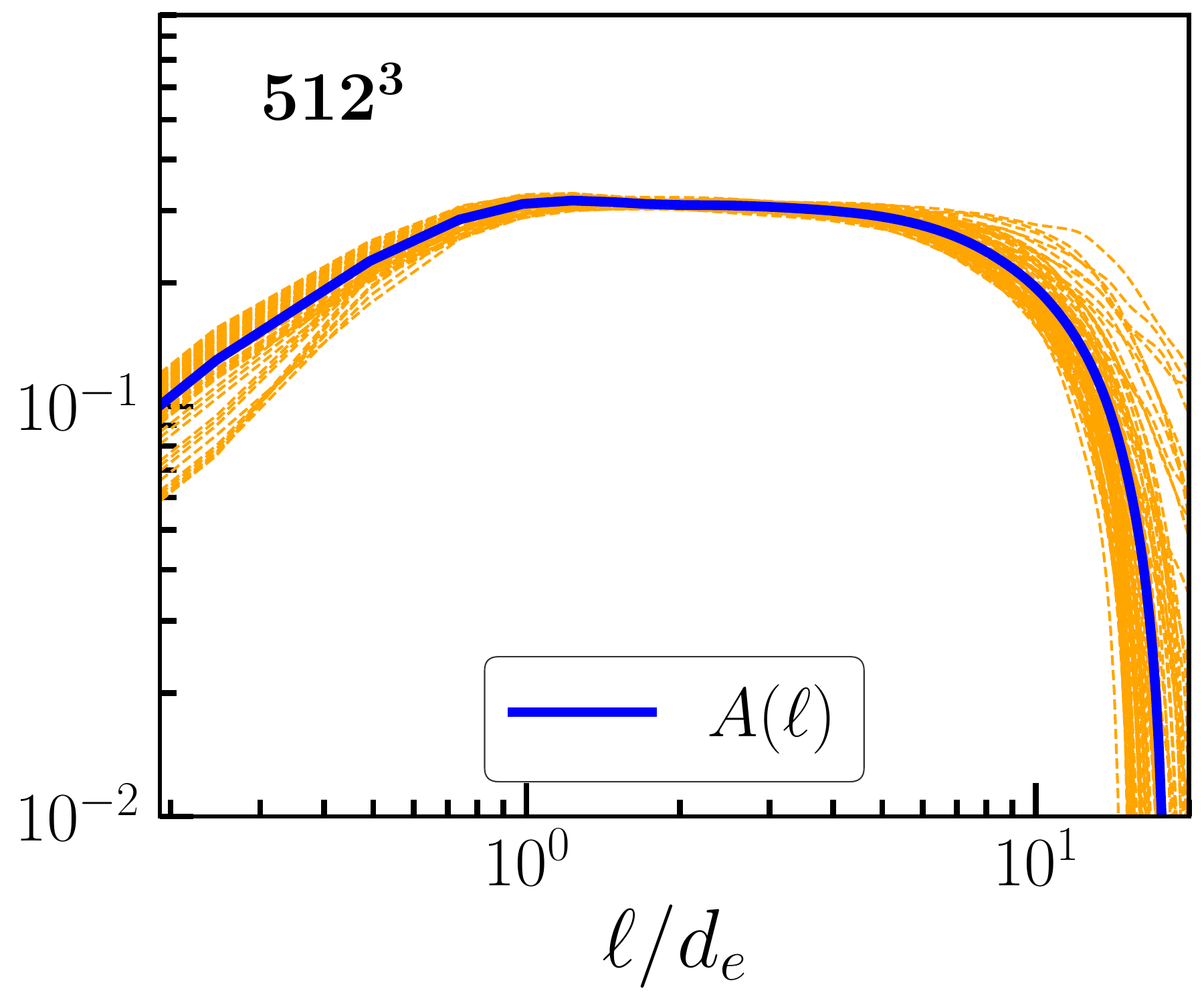}
    \caption{Total energy cascade rate as a function of normalised length scale. The dashed lines denote the cascade rates for all 73 directions whereas the solid line denotes their average for the simulation with $512^3$ grid points. }
    \label{exact_law_all_directions}
\end{figure}
In Fig.~\ref{exact_law_all_directions}, we plot the total energy cascade rates for individual directions (dashed lines) along with their average cascade rate (solid line) as a function of normalised length scale $\ell/d_e$. Interestingly, in an intermediate range of scales, the individual cascade rates become fairly identical and equals to the average cascade rate $A (\ell)$. This is possibly because the non-divergence form of exact relation is robust against the variation of directions across the aforesaid intermediate range of scales.

\begin{figure}
\centering
\includegraphics[width= \columnwidth]
{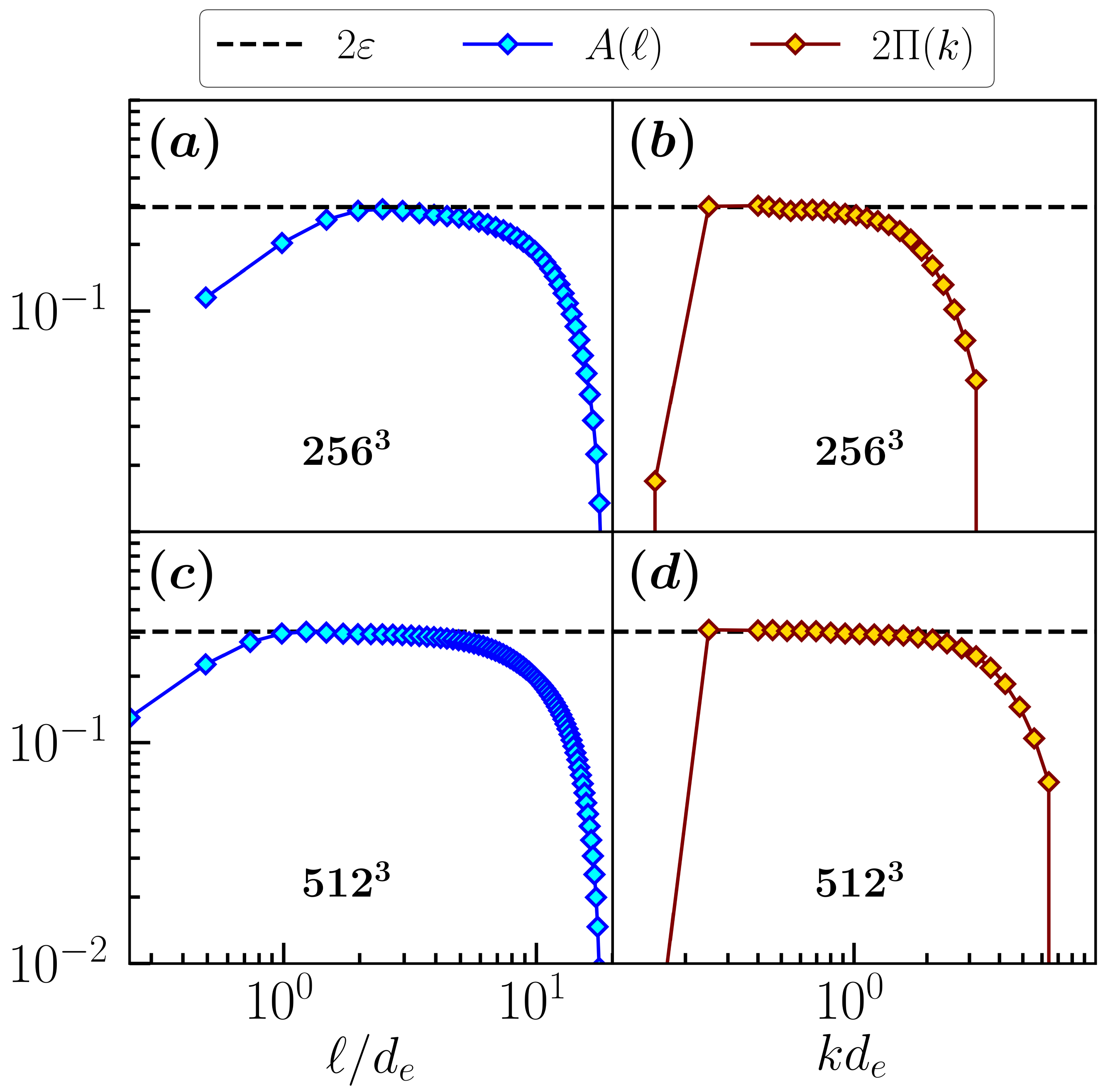}
    \caption{Total energy cascade rate as a function of normalised length-scale (wavenumber) for two grid resolutions. $A (\ell)$ and $\Pi(k)$ denote the scale-to-scale transfer rates obtained using the exact law and flux relation in Fourier space respectively. In all cases, a balance between average injection and cascade rate is observed inside the inertial range indicating a universal regime. A clear increase in the size of the inertial range is observed from resolution $256^3$ to $512^3$.}
    \label{flux_rate_real_and_spectral}
\end{figure}

Figs.~\ref{flux_rate_real_and_spectral} (a) and (c) show the energy cascade rates as a function of normalised length scale for runs with $256^3$ and $512^3$ grid points, respectively. For both cases, a flat region is found within the intermediate range of length-scales indicating a scale-independence of $A(\ell)$. Further, in accordance with statistical stationarity, this equals to twice the large-scale injection rate $\varepsilon$ (dashed line in Fig.~\ref{flux_rate_real_and_spectral}). These facts convincingly show the existence of a Kolmogorov-like universal energy cascade in 3D EMHD turbulence. Also as expected, a wider inertial zone is observed in the simulation with $512^3$ grid points.
To calculate the isotropic spectral flux rates in Eq.~\eqref{flux_Fourier_energy_nondim}, we adopt a more straightforward method than the one employed for structure function calculation. Here, the Fourier space is covered with 25 concentric wavenumber spheres of increasing radii: $k_i\in \{1, 3, 5, 8,..., N/4, N/2\}$. The radii between $8$ and $N/4$ is logarithmically binned with the ratio $k_{i+1}/k_i\approx 1.11$ and $1.15$ for run A and B, respectively. Such a choice of wavenumbers is found to be sufficient to capture the inertial range energy cascade. 
In Figs.~\ref{flux_rate_real_and_spectral} (b) and (d), we plot $\Pi (k)$ as a function of normalised wavenumbers $(kd_e)$ for resolutions $256^3$ and $512^3$, respectively. For both the cases, again a flat region is observed where the $\Pi(k)$ is constant and matches reasonably well with the injection rate (dashed black lines), thereby confirming the existence of a universal energy cascade. As in the previous case, a wider inertial range is clearly observed with the increase of grid points from $256^3$ to $512^3$. 
\begin{figure}
    \centering
    \includegraphics[width= 0.9\columnwidth]{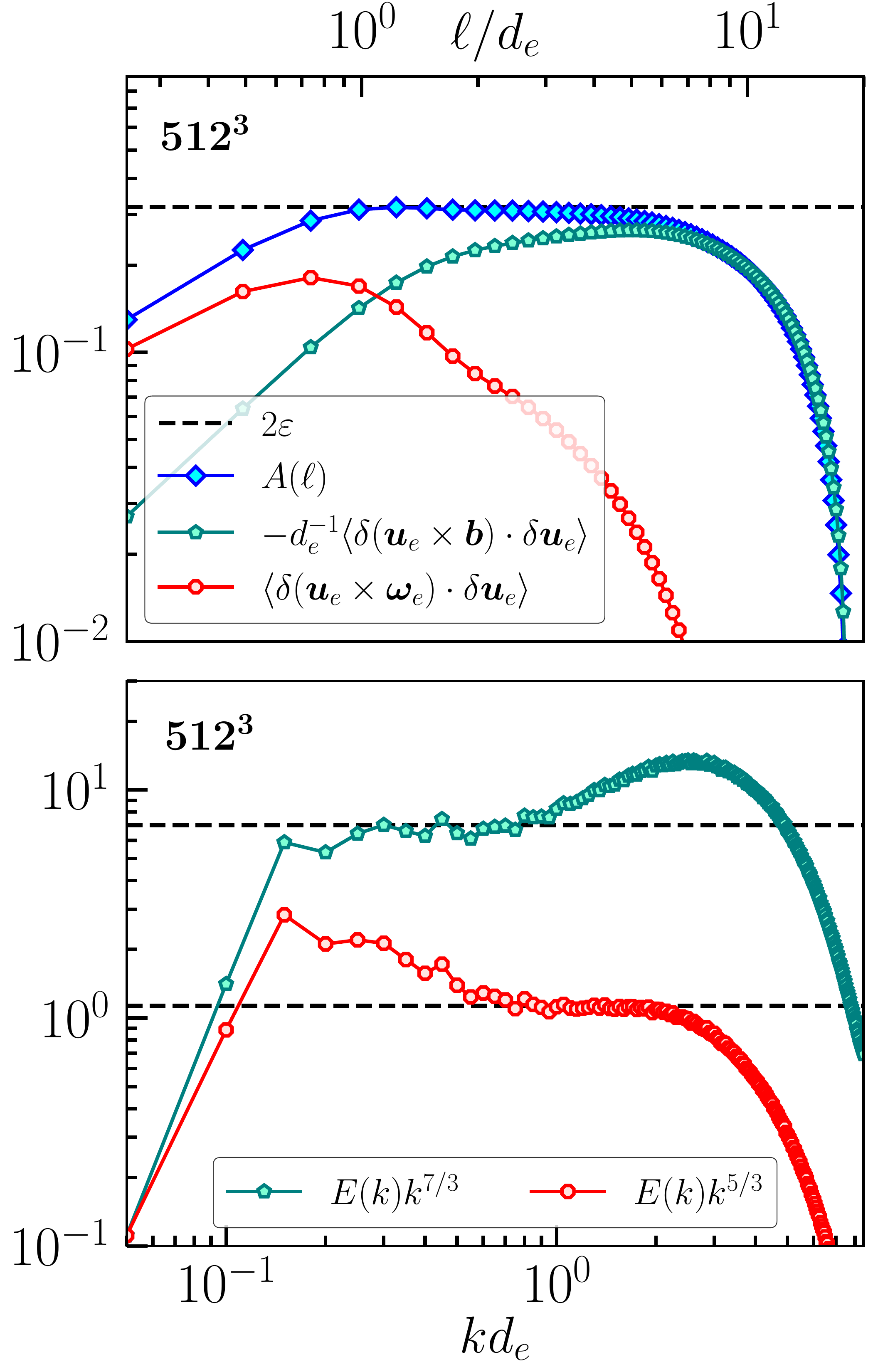}
    \caption{Total energy cascade rate as a function of normalised length-scale (top) and compensated energy spectra (bottom). Evidently, the magnetic energy cascade rate dominates for $\ell/d_e\gg 1$ (green line in top figure) corresponding to a $k^{-7/3}$ spectra for total energy (green line in bottom figure). The electron fluid kinetic energy cascade rate (red line in top figure) overtakes magnetic energy cascade rate at $\ell\simeq d_e$ and ultimately dominates it for $\ell/d_e\ll 1$. This is associated with the development of $k^{-5/3}$ spectra in the bottom figure (red line).  }
    \label{combined_detailed_and_spectra}
\end{figure}

To get a more detailed picture of the energy cascade rate, using the relation $\bm{q} = \bb - d_e\bw_e$, we decompose Eq.~\eqref{exact_energy_nondim} as 
\begin{equation}
   A(\bm{\ell})= \left\langle\delta(\bu_e\times\bw_e)\cdot\delta\bu_e\right\rangle -\frac{1}{d_e}\left\langle\delta(\bu_e\times\bb)\cdot\delta\bu_e\right\rangle = 2\varepsilon.
\end{equation}
For $\ell\gg d_e$, where the inertia of the electron fluid is negligible and the dynamics is primarily governed by the magnetic field, the first term on the \textit{l.h.s.} can be neglected with respect to the second term. On the other hand, for small scales $\ell\ll d_e$, the inertia of the electron fluid becomes non-negligible and 
the resultant energy cascade rate is practically given by that of the electron fluid kinetic energy. In Fig.~\ref{combined_detailed_and_spectra} (top), we plot these two components of the exact relation separately as a function of normalised length scale. The blue and red solid curves respectively denote the cascade rates of magnetic and electron fluid kinetic energy. As expected, for large scales $\ell/d_e\gg 1$, the magnetic energy cascade rate dominates over the kinetic cascade rate. From a phenomenological perspective, this regime corresponds to a $k^{-7/3}$ spectra for total energy as shown in Fig.~\ref{combined_detailed_and_spectra} (bottom). For smaller scales, the electron fluid inertia gradually increases thus developing a prominent kinetic energy cascade for the electron fluid. At $\ell/d_e\simeq 1$, the kinetic energy cascade rate catches up the cascade rate for magnetic energy and eventually surpasses it thereafter. This marks the inertial EMHD regime characterized by the emergence of a Kolmogorov-type $k^{-5/3}$ spectra for the electron fluid kinetic energy cascade (see Fig.~\ref{combined_detailed_and_spectra} bottom).

\begin{figure}
    \centering
    \includegraphics[width=0.9\columnwidth]{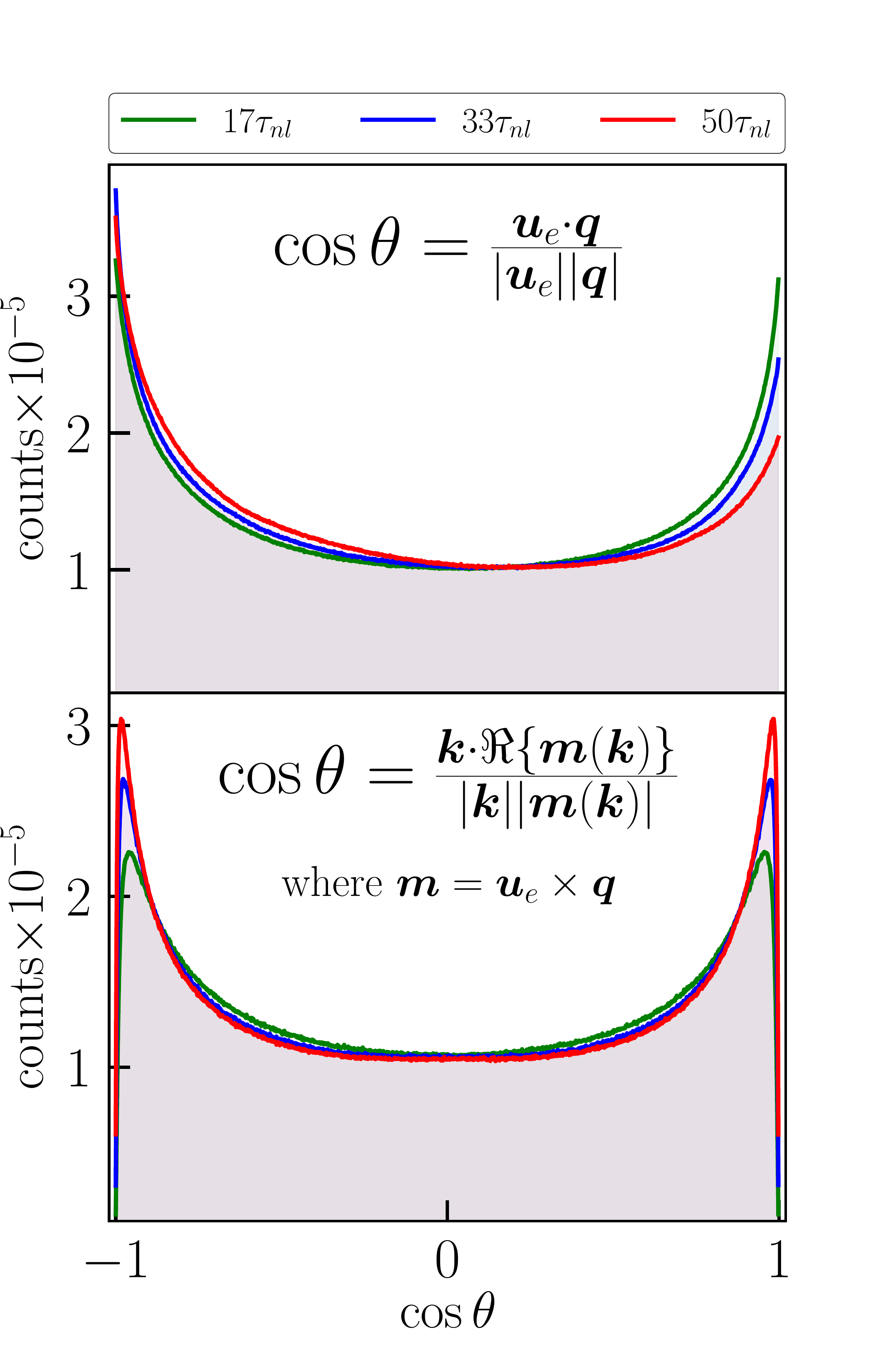}
    \caption{Histograms of $\cos\theta$ where $\theta$ is the angle between the vectors $\bu_e - \bm{q}$ (top) and $\bk - \Re\{\bm{m}(\bk)\}$ (bottom) where $\bm{m}(\bk) = \widehat{\left(\bu_e\times\bm{q}\right)}_{\bk}$ vs. time. For the second case, a clear broadening of the histograms towards $\pm 1$ with time confirms an alignment between $\bk$ and $\bm{m}(\bm{k})$ thereby validating Eq.~\eqref{relaxed_state_nondim}.}
    \label{rlx_two_angles}
\end{figure}

The next objective of this paper is to investigate and characterize the states where a fully developed EMHD turbulent flow would relax upon the withdrawal of the turbulent drive. To achieve that,  we quench the forcing of run B at $16\tau_{nl}$ and let the system undergo turbulent relaxation. 
\begin{figure}
    \centering
    \includegraphics[width= 0.9\columnwidth]{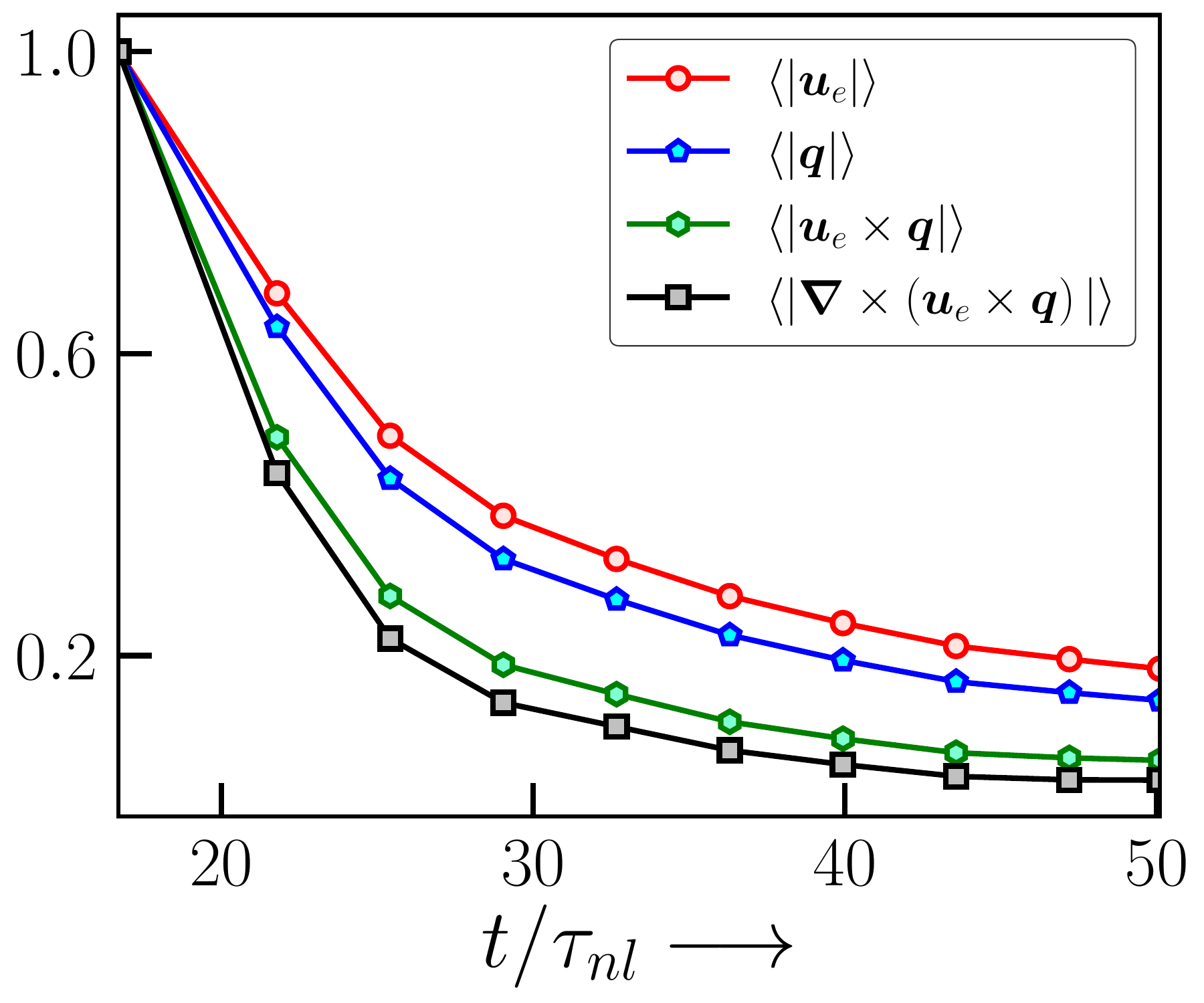}
    \caption{Average values of various dynamical variables as function of normalised time. As expected for a relaxation process, average values of all the dynamical variables diminishes with time. Clearly, $\langle|\bnabla\times\left(\bu_e\times\bm{q}\right)|\rangle$ decays fastest, implying the possibility of a pressure-balanced relaxed state.} 
    \label{avg_nlin_vs_time}
\end{figure}
According to PVNLT, the relaxed states are obtained when the average nonlinear transfers corresponding to the cascade of inviscid invariants vanish. For an EMHD fluid, a non-trivial relaxed state ($\bu_e\neq \bm{0},\  \bm{q}\neq \bm{0}$) is therefore obtained when  $\bnabla\times\left(\bu_e\times\bm{q}\right)$ vanishes. 
Fig.~\ref{avg_nlin_vs_time}, shows the evolution of the average values of $\langle|\bu_e|\rangle$, $\langle|\bm{q}|\rangle$, $\langle|\bu_e\times\bm{q}|\rangle$ and $\langle|\bnabla\times\left(\bu_e\times\bm{q}\right)|\rangle$ as a function of time (normalised by $\tau_{nl}$). Veritably, the average nonlinear terms $\langle|\bu_e\times\bm{q}|\rangle$ and $\langle|\bnabla\times\left(\bu_e\times\bm{q}\right)|\rangle$ fall off much quicker than $\langle|\bu_e|\rangle$ and $\langle|\bm{q}|\rangle$. Interestingly, of the two nonlinear terms, $\langle|\bnabla\times\left(\bu_e\times\bm{q}\right)|\rangle$ decreases faster than $\langle|\bu_e\times\bm{q}|\rangle$. This indicates the possibility of a relaxed state where $\bu_e\times\bm{q}$ becomes irrotational while remaining a non-zero vector. In order to further endorse this possibility, we plot the histogram of the cosine of angle $(\theta)$ between $\bu_e$ and $\bm{q}$ for three different times $17\tau_{nl}$, $33\tau_{nl}$ and $50\tau_{nl}$ (Fig.~\ref{rlx_two_angles} top).
As time increases, the number of the points associating $\bu_e-\bm{q}$ alignment (or anti-alignment) does not increase unambiguously. However, the number of points with vanishing $\bnabla\times(\bu_e\times\bm{q})$ increases unambiguously as the relaxation continues. This can directly be verified from the gradually increasing alignment (or anti-alignment) between the vectors $\bk$ and the real part of $\widehat{\left(\bu_e\times\bm{q}\right)}_{\bk}$ in the Fourier space (see Fig.~\ref{rlx_two_angles} bottom). Similar behaviour is observed in the histograms for the cosine of the angle between $\bk$ and imaginary part of $\widehat{\left(\bu_e\times\bm{q}\right)}_{\bk}$ (not shown here). Rather than a $\bu_e-\bm{q}$ alignment,  our findings indicate the existence of a relaxed state where the 
vector $\bu_e\times\bm{q}$ reduces to the gradient of a scalar function.

\section{Discussion}
\label{sec4}
In this paper, we derive a divergence-free exact relation 
for the total energy cascade in three-dimensional homogeneous EMHD turbulence. Owing to the divergence-free form, the exact relation is equally valid for isotropic as well as non-isotropic turbulent flows. The exact relation for energy cascade is particularly important because it provides an accurate measure of the heating rate due to the nonlinear cascades for length-scales comparable or smaller than $d_e$. In the limit of length-scales much larger and smaller than $d_e$, as expected, the exact relation \eqref{exact_law_energy_final} practically correspond to the inertia-less and inertial dominated EMHD regimes, respectively. For $kd_e\ll 1$, the total energy cascade essentially represents a magnetic energy cascade whereas for $kd_e\gg 1$, it represents a kinetic energy cascade for electron fluid. By explicit calculation, we have also showed that the total energy cascade rate is unaffected by the presence of a uniform background magnetic field $\bB_0$. Complementary to the exact relation in direct space, the universal energy cascade rate is also calculated using scale-by-scale energy transfer rate in the Fourier space (see Eq.~\eqref{Flux_rate_energy_Fourier}). Finally, using the recently proposed PVNLT, we analytically predicted the turbulent relaxed states where the turbulent cascades disappear. For three-dimensional EMHD flow, the corresponding relaxed states are characterized by $\bnabla\times(\bu_e\times\bf{Q}) = \bm{0}$. 


Using direct numerical simulations with up to $512^3$ grid points, we convincingly showed the existence of a Kolmogorov-type energy cascade associating a constant energy flux rate across the inertial scales. At stationary state, the flux rate $A(\ell)$ is found to become constant across scales and equal to twice the one-point average energy injection rate $\varepsilon$ which, in turn, is also equal to the average heating rate due to EMHD cascade. Similarly, in Fourier space, the scale-independent cascade rate was obtained by calculating average energy flux rate $\Pi (k)$. As expected for a Kolmogorov cascade, the inertial range becomes wider for the high resolution run (B) leading to a smaller Kolmogorov scale. A term-by-term analysis of the exact relation \eqref{exact_energy_nondim} clearly shows a contrast between the flux rates of magnetic and electron fluid kinetic energies, respectively. The magnetic energy cascade rate dominates for scales larger than $d_e$ whereas the kinetic energy transfer rate takes over for scales inferior to $d_e$. These two sub-ranges are phenomenologically characterized as the inertia-less and inertia-dominated regimes corresponding to a Hall-like $k^{-7/3}$ and hydrodynamic-like $k^{-5/3}$ power spectra, respectively (see Fig. \ref{combined_detailed_and_spectra}).   

Finally, the turbulent drive was quenched, allowing the system to attain the relaxed states. During the process of relaxation, the curl of the  vector $\bu_e \times \bm{q}$ was clearly found to quickly fall off thereby making a perfect agreement with PVNLT (see Fig.~\ref{avg_nlin_vs_time}). However, as shown by our simulation results, this behaviour does not simply emerge from a mere alignment (or anti-alignment) between $\bu_e$ and $\bm{q}$ but is associated to a pressure-balanced relaxed state instead (see Fig.~\ref{rlx_two_angles}). 

The current work is a direct evidence of Kolmogorov-type energy cascade in three-dimensional inertial EMHD turbulence. The derived exact law enables precise measurement of the turbulent heating rate in the solar wind using high resolution data of MMS (NASA), Parker solar probe (NASA), Solar orbiter (ESA), and other missions. In addition, the cascade rate can be pivotal in understanding how an EMHD turbulence influences the magnetic reconnection rates in space plasmas. Employing a helical forcing, a separate work can be dedicated to the nature (direct or inverse) and universality of the stationary helicity cascade in EMHD turbulence. Again, one can derive an exact relation for reduced EMHD incorporating compressibility in the presence of a strong background magnetic field. Such a model can be crucial for understanding certain aspects of kinetic Alfv\'en wave turbulence, a key framework for understanding many small-scale phenomena in space and astrophysical plasma turbulence.   

\appendix
\section{Conservation and exact law for canonical helicity}
\label{appA}
In the absence of $\bm{f}_{\bq}$ and $\bm{D}_{\bf Q}$, Eq.~\eqref{frozen_in_Eq} reduces to  
\begin{equation}
   \frac{\partial {\bf Q}}{\partial t} = \bnabla \times \left( \bu_e \times {\bf Q} \right),  \label{frozen_in_Eq_ideal} 
\end{equation}
which clearly signifies that the field ${\bf Q}$ is frozen in the electron fluid. Uncurling Eq.~\eqref{frozen_in_Eq_ideal} one can also get
\begin{equation}
   \frac{\partial {\bf P}}{\partial t} =  \bu_e \times {\bf Q} + \bnabla \psi \label{gen_vec_potential}  
\end{equation}
where $\bP$ is the vector potential of ${\bf Q}$ and $\psi$ is a scalar gauge function. From Eq.~\eqref{emhd_Faraday}, in general, one can write $\bE = -\partial\bA/ \partial t - \bnabla\phi$ where $\phi$ and $\bA$ are the electrostatic potential and magnetic vector potential, respectively. Substituting the expression of $\bE$ in Eq.~\eqref{emhd_mom}, it is straightforward to show ${\bf P} = \bA - (m_e\bu_e)/e$ and $\psi = p_e/(ne) - \phi $. Using the structure of Eqs.~\eqref{frozen_in_Eq} and \eqref{frozen_in_Eq_ideal} and the proportionality between $\bu_e$ and $\bJ$, one can show that 
\begin{equation}
\frac{\partial \langle H\rangle}{\partial t}  = \langle\bnabla\cdot\left[\left(\bu_e\times{\bf Q}\right)\times\bP +\psi\bf{Q}  \right]\rangle, 
\end{equation}
where $H = \bP \cdot {\bf Q}$ is the canonical helicity density. Again assuming periodic or vanishing boundary conditions, one can show that $\langle H\rangle$ is an inviscid invariant of the system.

The symmetric two-point correlator for $\langle H\rangle$ is given by  
\begin{align}     
     R_H &= \left\langle \frac{\bP \cdot \bq' + \bP' \cdot \bq}{2} \right\rangle .
\end{align}
Using Eqs.~\eqref{frozen_in_Eq_ideal} and \eqref{gen_vec_potential}, the corresponding evolution is given by 
\begin{align}
    &\frac{\partial R_H}{\partial t}= \frac{1}{2} \left\langle \frac{\partial \bP}{\partial t} \cdot \bq' +  \frac{\partial \bP'}{\partial t} \cdot \bq + \frac{\partial \bq}{\partial t} \cdot \bP' + \frac{\partial \bq'}{\partial t} \cdot \bP   \right\rangle \nonumber\\
    &= \frac{1}{2} \left\langle \left( \bu_e \times {\bf Q} + \bnabla \psi\right) \cdot \bq' + \left(\bu'_e \times {\bf Q'} + \bnabla' \psi'\right) \cdot \bq \right.\nonumber \\
    &+ \left. \bnabla \times \left( \bu_e \times {\bf Q} \right)  \cdot \bP' +  \bnabla' \times \left( \bu'_e \times {\bf Q'} \right) \cdot \bP \right\rangle + {\cal D}_H + {\cal F}_H \nonumber \\
    &= \left\langle \left( \bu_e \times {\bf Q} \right) \cdot \bq' + \left(\bu'_e \times {\bf Q'} \right) \cdot \bq \right\rangle + {\cal D}_H + {\cal F}_H, \label{time_derivative_correlator_final_H}
\end{align}
where ${\cal F}_H$ and ${\cal D}_H$ represent the average two-point injection and dissipation terms for $\langle H\rangle$, respectively.  Again (i) assuming a statistical stationary state for the average helicity transfer, (ii) neglecting the dissipative effects for length scales far from the dissipative scales, and (iii) equating ${\cal F}_H (\bm{0}) = \varepsilon_H$, we get 
\begin{equation}
A_H(\bm{\ell}) = \left\langle \delta \left( \bu_e \times {\bf Q} \right) \cdot \delta \bq  \right\rangle =  \varepsilon_H, \label{exact_law_final_helicity}
\end{equation}
where $A_H(\bm{\ell})$ is the average transfer rate of $\langle H\rangle$ across scales.

Similar to the case of total energy, one should not attempt for a `hydrodynamic limit' simply by putting $\bB = \bm{0}$ in Eq.~\eqref{exact_law_final_helicity}. As before, for length scales much larger and smaller than $d_e$, one can effectively write 
\begin{align}
\left\langle\delta(\bu_e\times\bB)\cdot\delta\bB\right\rangle &= \varepsilon_H, \;\; \text{for}\;\; kd_e\ll 1\;\; \text{and}\label{exact_law_helicity_large}\\
\left(\frac{m_e}{e}\right)^2\left\langle\delta(\bu_e\times\bw_e)\cdot\delta\bw_e\right\rangle &= \varepsilon_H, \;\; \text{for}\;\; kd_e\gg 1,\label{exact_law_helicity_smaller}
\end{align}
respectively. Obviously, at large scales $kd_e\ll 1$, Eq.~\eqref{exact_law_helicity_large} resembles the exact relation for magnetic helicity cascade in inertia-less EMHD whereas at very small scales $kd_e\gg1$, Eq.~\eqref{exact_law_helicity_smaller} effectively represents the kinetic helicity cascade for the electron fluid across the scales dominated by electron inertia \citep{Banerjee_2016a, Banerjee_2016b}.

Unlike $\varepsilon$, one can show that $\bB_0$ gives a non-zero contribution in $\varepsilon_H$ which is given by
$\left\langle(\delta\bu_e\times\bB_0)\cdot\delta\bq\right\rangle$. A careful inspection further shows that this nonzero contribution of $\bB_0$ in $\varepsilon_H$ comes due to the large scale magnetic helicity cascade given by the Eq.~\eqref{exact_law_helicity_large} while the small scale kinetic helicity cascade rate in Eq.~\eqref{exact_law_helicity_smaller} remains unaltered in the presence of $\bB_0$.


\begin{acknowledgments}
S.B. and A.H. acknowledge the financial support from STC-ISRO grant (STC/PHY/2023664O). A.H. acknowledges N. Pan for helping with the parallelization of structure function calculation code. The simulation code is developed by AH following the parallelization schemes in Ref. \cite{Mortensen_2016}. The simulations are performed using the support and resources provided by PARAM Sanganak under the National Supercomputing Mission, Government of India at the Indian Institute of Technology, Kanpur for providing computational resources. 
\end{acknowledgments}

\section*{author declarations}
\subsection*{Conflict of interest}
The authors declare no conflicts of interest.
\subsection*{Data availability} 
The simulation data will be made available upon reasonable request to the corresponding author.
\section*{author contributions}
Supratik Banerjee and Arijit Halder equally contributed to the paper.

\noindent\textbf{Supratik Banerjee:} Conceptualization (equal); Formal analysis (lead); Validation (supporting); Visualization (supporting);
Writing - original draft (equal); Writing - review and editing (supporting).  \\
\textbf{Arijit Halder:} Conceptualization (supporting); Formal analysis (supporting); Validation (lead); Visualization (lead); 
Writing - original draft (equal); Writing - review and editing (supporting).\\
\textbf{Amita Das:} Conceptualization (equal); Formal analysis (supporting); Validation (supporting); Visualization (supporting); Writing - review and editing (lead).

%

\end{document}